\documentclass[apj]{emulateapj}

\newcommand{\simgt}{\,\hbox{\lower0.6ex\hbox{$\sim$}\llap{\raise0.6ex\hbox{$>$}}}\,}

\newcommand{\simlt}{\,\hbox{\lower0.6ex\hbox{$\sim$}\llap{\raise0.6ex\hbox{$<$}}}\,}

\shorttitle{A Line-Of-Sight Cluster Collision}
\shortauthors{ZuHone et al.}
        
\begin{document}

\title{A Line-Of-Sight Galaxy Cluster Collision: \\
Simulated X-Ray Observations}

\author{J. A. ZuHone}
\affil{Department of Astronomy and Astrophysics, University of Chicago, Chicago, IL 60637}
\email{zuhone@flash.uchicago.edu}
\and
\author{P. M. Ricker}
\affil{Department of Astronomy, University of Illinois at Urbana-Champaign, Urbana, IL, 61801}
\email{pmricker@uiuc.edu}
\and
\author{D. Q. Lamb}
\affil{Department of Astronomy and Astrophysics, University of Chicago, Chicago, IL 60637}
\email{lamb@oddjob.uchicago.edu}
\and
\author{H.-Y. Karen Yang}
\affil{Department of Astronomy, University of Illinois at Urbana-Champaign, Urbana, IL, 61801}
\email{hyang20@illinois.edu}

\begin{abstract}
Several lines of evidence have suggested that the the galaxy cluster Cl~0024+17, an apparently relaxed system, is actually a collision of two clusters, the interaction occurring along our line of sight. In this paper we present a high-resolution $N$-body/hydrodynamics simulation of such a collision. We have created mock X-ray observations of our simulated system using MARX, a program that simulates the on-orbit performance of the Chandra X-ray Observatory. We analyze these simulated data to generate radial profiles of the surface brightness and temperature. At later times, $t = 2.0-3.0$ Gyr after the collision, the simulated surface brightness profiles are better fit by a superposition of two $\beta$-model profiles than a single profile, in agreement with the observations of Cl~0024+17. In general, due to projection effects, much of the post-collision density and temperature structure of the clusters is not seen in the observations. In particular, the observed temperatures from spectral fitting are much lower than the temperature of the hottest gas. We determine from our fitted profiles that if the system is modeled as a single cluster, the hydrostatic mass estimate is a factor $\sim$2-3 less than the actual mass, but if the system is modeled as two galaxy clusters in superposition, a hydrostatic mass estimation can be made which is accurate to within $\sim$10$\%$. We examine some implications of these results for galaxy cluster X-ray surveys.
\end{abstract}


\section{Introduction}

Clusters of galaxies have proven to be interesting laboratories for studying the dynamics of the different kinds of matter in the universe. Early measurements of galaxy velocity dispersions in clusters demonstrated that most of the mass of galaxy clusters was in the form of a non-luminous component \citep{zwy37}. Modern cosmological constraints (e.g. Turner 2000) suggest that this non-luminous material must be mostly non-baryonic. In addition, X-ray observations of clusters of galaxies reveal that most of their baryonic mass is in the form of a hot, diffuse, X-ray emitting gas (the intracluster medium or ICM). The presence of these different types of matter (stellar, gaseous, and dark) in these systems provides not only insights into the nature of the clusters themselves but also the nature of the different components of matter and their dynamics. 

In the cold dark matter (CDM) picture of structure formation, galaxy clusters are the product of many mergers between galaxies, galaxy groups, and smaller clusters of galaxies \citep{dav85}. Observations of galaxy clusters indicate that this process is still ongoing in many systems. A famous example is the so-called ``Bullet Cluster'' (1E~0657-56), a merging system in which X-ray and weak gravitational lensing observations demonstrate a clear separation between a collisionless dark matter component and the intracluster medium \citep{mar02, clo06}. 

Another system that has recently attracted attention is Cl~0024+17, a cluster at a redshift $z = 0.395$ exhibiting both weak and strong lensing. Early attempts at reconstructing the mass profile of this system using lensing \citep{tys98, bro00, com06} suggested a conflict with the predictions of the standard CDM model due to the flattening of the density profile in the inner regions of the cluster. Cosmological simulations assuming CDM indicate that galaxy cluster mass profiles should exhibit a nonzero logarithmic slope in the inner regions (e.g. Navarro, Frenk, \& White 1997). This apparent discrepancy (and others) led to suggestions that the CDM paradigm would need to be modified \citep{spr00,hog00,moo00}, for example to include self-interaction of the dark matter. 

\citet{czo01, czo02} demonstrated that the redshift distribution of the cluster galaxies in Cl~0024+17 is bimodal, with a large, primary component and a more diffuse, secondary component, separated from the primary component in velocity space by $v \sim$ 3000 km/s. They suggested that the system is composed of two clusters undergoing a high-velocity collision along the line of sight. They also performed a simulation demonstrating that such a scenario reproduces not only the bimodal redshift distribution but also the flattening of the central density profile. X-ray observations of the cluster using Chandra \citep{ota04} and XMM-Newton \citep{zhg05} revealed that the surface brightness profile is better fit by a superposition of two ICM models rather than one. They suggested on the basis of the isothermal temperature profile that the individual clusters of the system had returned to equilibrium after the collision and that consequently the encounter must have occurred several Gyr ago.

Recently, a weak lensing analysis presented by \citet{jee07} revealed a ringlike structure in the projected matter distribution. They proposed that dark matter from the cores of the clusters had been disrupted and ejected from the systems by the collision, and they demonstrated that such features could be reproduced in a simulation of a collision of two pure dark matter halos. From this they suggested the current state of the system is $\sim$1-2 Gyr past the pericentric passage of the cluster cores. We are addressing this issue in a separate paper (ZuHone et al. 2008, in preparation). \citet{jee07} also demonstrated that if the system is modeled as two ICM profiles in superposition, the mass determination based on hydrostatic equilibrium agrees with the result from their lensing analysis. 

If the merger scenario for Cl~0024+17 is correct, it raises several important questions: how long after the collision would the clusters appear relaxed, and correspondingly when would a mass estimate based on hydrostatic equilibrium yield an accurate measurement? What effect does viewing such a collision along the line of sight, with both cluster components in superposition, have on the observed density and temperature structure? In this paper we seek to provide answers to these questions by simulating a similar high-speed collision between two clusters of galaxies. Most importantly, we include the dynamics of the cluster gas, which were not included in the previous simulations.

Throughout this paper we assume a flat CDM cosmology with $h = 0.5$, $\Omega_{\rm m} = 1.0$, as in \citet{ota04}. 

\section{Simulations}

\subsection{Method}

We performed our simulations using FLASH, a parallel hydrodynamics/$N$-body astrophysical simulation code developed at the Center for Astrophysical Thermonuclear Flashes at the University of Chicago \citep{fry00}. FLASH uses adaptive mesh refinement (AMR), a technique that places higher resolution elements of the grid only where they are needed. In our case we are interested in capturing sharp ICM features like shocks and ``cold fronts'' accurately, as well as resolving the inner cores of the cluster dark matter halos. As such it is particularly important to be able to resolve the grid adequately in these regions. AMR allows us to do so without needing to have the whole grid at the same resolution. FLASH solves the Euler equations of hydrodynamics using the Piecewise-Parabolic Method (PPM) of \citet{col84}, which is ideally suited for capturing shocks. FLASH also includes an $N$-body module which uses the particle-mesh method to solve for the forces on gravitating particles. The gravitational potential is computed using a multigrid solver included with FLASH \citep{ric08}.

\subsection{Initial Conditions}

The cluster dark matter halos are initialized using the NFW \citep{NFW97} density profile, where we follow the method of \citet{kaz06}. The NFW functional form is used for $r \leq r_{200}$:
\begin{equation}
\rho_{\rm DM}(r) = {\rho_{\rm s} \over {r/r_{\rm s}{(1+r/r_{\rm s})}^2}},\ r \leq r_{200}\ .
\end{equation}
Outside of $r_{200}$, to keep the mass of the halo finite and to avoid an unphysical distribution function, we implement an exponential cutoff that turns off the profile on a scale $r_{\rm decay}$, a free parameter which we set to $0.1r_{200}$:
\begin{eqnarray}
\lefteqn{\rho_{\rm DM}(r) = {\rho_{\rm s} \over
    {c_{200}(1+c_{200})^2}}{\left({r \over r_{200}}\right)^\kappa}}
\nonumber\\ & & \times {}{\rm exp}{\left(-{r-r_{200} \over r_{\rm decay}}\right)},\ r > r_{200}\ .
\end{eqnarray}
Here $r_{200}$ is the radius within which the mean mass density is 200 times the cosmic mean, $r_{\rm s}$ is the NFW scale radius, and $c_{200} \equiv r_{200}/r_{\rm s}$ is the NFW concentration parameter. We require that at $r_{200}$ the profile and its logarithmic slope be continuous. This is achieved via the parameter $\kappa$:
\begin{equation}
\kappa = -{(1+3c_{200}) \over (1+c_{200})} + {r_{200} \over r_{\rm decay}}  
\end{equation}
Truncating the cluster NFW profiles at $r_{200}$ is unphysical.  In real dark matter halos, the density profile is expected to remain NFW-like out to where the density of the halo reaches the average matter density of the universe \citep{pra06,tav08}.  Single cluster tests we have performed (as well as those performed in other studies, e.g., Ricker and Sarazin 2001) demonstrate that the truncated density profile is stable for the length of the run, with minor steepening of the profile near the cutoff radius.

The truncation also reduces the halo mass by approximately 25$\%$.  Assuming the clusters fell in from infinity, the larger halo masses would accelerate the halos to an impact velocity approximately 12$\%$ larger than the value we use.  After the collision, the corresponding deceleration due to gravity would also be higher.  The effect of the larger halo masses on the deceleration due to dynamical friction ($f_{\rm dyn} \propto \rho{M^2}/v^2$) is more complicated to determine, since although the density in the outskirts of the halos is higher (thus increasing the dynamical friction), the velocity of clusters would also be higher due to the increased acceleration (thus decreasing it). The higher infall velocity of the clusters and the increased mass outside $r_{200}$ might also lead to some increase in the dissipation of the shock energy, and therefore in the heating of the gas in the outskirts.  However, in general any such effects are inaccessible to current X-ray observations, due to the low densities in the cluster outskirts. We conclude that the effects due to truncation of the cluster density profile are not likely to be significant for the purposes of this study.

To initialize the gas density we choose a Burkert profile:
\begin{equation}
\rho_{\rm gas}(r) = {\rho_{\rm c} \over {(1 + ({r/r_{\rm c}})^2)(1 + {r/r_{\rm c}})}}\ ,
\end{equation}
which originally \citep{bur95} was given as a fitting function for
dark matter profiles of dwarf galaxies but is also a good fit to the
gas density profiles of non-cooling-core clusters (A. Kravtsov 2007, private communication). The more traditional $\beta$-model \citep{cav76} is a poorer fit to the gas density profile at large $r$ in both simulations and observations \citep{vik06,hal07,nag07}. (However, we follow \citet{ota04} and \citet{zhg05} and fit $\beta$-model profiles to the X-ray surface brightness profiles in our simulated observations.)

The gas density is also fitted to an exponential taper at $r_{200}$, extending to the radius at which it equals the mean baryonic density of the universe. For the gas profiles, we choose the core radius $r_{\rm c}$ to be roughly half the scale radius of the DM profile \citep{ric01}. We fix the gas mass fraction at $r_{200}$ to 0.12 in line with observations of real clusters (e.g., Vikhlinin et al. 2006, Sanderson et al. 2003), and the measurement of \citet{ota04}.

From these parameters the central gas densities $\rho_{\rm c}$ can be determined. The equation of state for the gas is an ideal-gas equation of state with $\gamma = 5/3$ and mean atomic mass $\mu$ = 0.592. We determine the pressure, temperature, and internal energy profiles by assuming our functional forms for dark matter and gas density and numerically integrating the equation of hydrostatic equilibrium. The values used for the cluster parameters are given in Table 1. 

\begin{deluxetable}{cccccc}
\tabletypesize{\scriptsize}
\tablecaption{Initial Cluster Parameters\label{tab1}}
\tablewidth{0pt}
\tablehead{
\colhead{Cluster} & \colhead{$M_{200}$ (M$_{\sun}$)} & 
\colhead{$r_{200}$ ({\rm kpc})} & \colhead{$c_{200}$} & 
\colhead{$n_{\rm c}$ (cm$^{-3}$)} & \colhead{$r_{\rm c}$ ({\rm kpc})}
}
\startdata
1 & $6.0 \times 10^{14}$ & 1739.79 & 5.0 & 0.020 & 198.83 \\
2 & $3.0 \times 10^{14}$ & 1380.87 & 7.0 & 0.062 & 98.65 \\
\enddata
\end{deluxetable}

Following \citet{czo02} and \citet{jee07} we assume a mass ratio of 2:1 for the clusters. In our simulation the clusters are initialized so that their centers are separated by the sum of their respective $r_{200}$ values (approximately 3 Mpc), and they are given an initial relative velocity $v_{\rm rel} = 3000$ km/s. This value is chosen because it is the inferred relative velocity of the two redshift components of Cl~0024+17 seen in the redshift histograms of  \citet{czo02}. The results of \citet{czo02} and \citet{jee07} suggest the clusters have already completed their first pericentric passage, and \citet{czo02} show that in their collision simulation an initial velocity of $\sim$3000 km s$^{-1}$ is needed to reproduce the observed distribution of redshifts after the collision. This value for the velocity is approximately 2$v_{ff}$ for the clusters, where $v_{ff}$ is the free-fall velocity from infinity. Such an initially high relative velocity might be difficult to achieve in a $\Lambda$CDM universe \citep[see, e.g.][]{hay06}. However, our choice is motivated by our desire to match the relative inferred velocity of the cluster components of Cl~0024+17.

We refine the adaptive mesh using the second derivative of density, temperature, and pressure to capture the sharp features in the gas dynamics, and using the matter density to resolve the cores of the clusters. For our box size of 14.29 Mpc we achieve a finest resolution of ${\Delta}x = 13.96$ kpc.

\subsection{Simulated Observations}

To make a meaningful comparison with the X-ray observations of Cl~0024+17, we construct mock X-ray observations of the simulation output. To do this we use MARX, a suite of programs created by the MIT/CXC group that simulates the on-orbit performance of the Chandra X-ray Observatory\footnote{http://space.mit.edu/ASC/MARX/}. MARX provides detailed ray-tracing simulations of astrophysical sources as observed by Chandra. In this paper we use MARX to generate simulated surface brightness profiles and spectra.

For input into MARX we create projected flux maps from the FLASH simulation data. For each FLASH zone the MEKAL model \citep{MKL95} is used to assign a spectral emissivity. For these simulations zero metallicity has been assumed. The spectra are projected along the appropriate line of sight, and flux maps are generated for photons in 198 separate energy bins from 0.1-10.0 keV, giving an energy resolution of ${\Delta}E$ = 50 eV, compared to the ${\Delta}E \approx$ 100 eV resolution of Chandra.
 
For our simulated Chandra observations we use the simulated ACIS-S detector and HRMA mirror system, as in the real Chandra observation of Cl~0024+17. Each observation is $\sim$40 ks of simulated exposure time, comparable to the Chandra observation of Cl~0024+17 (39 ks). The flux maps serve as a distribution function for the simulated photons and determine the direction cosines, energies, and timesteps for the photons in MARX. For simplicity, effects such as foreground galactic absorption and point source contamination are not included. However, we do include a diffuse X-ray background component. Output is given as a FITS file that can be read and analyzed in the same way as real Chandra observations. To analyze our simulated data, we use CIAO 3.4 and XSPEC 12.3. Details of our procedure for generating these observations and a verification study are given in the Appendix.


\begin{figure*}
\plotone{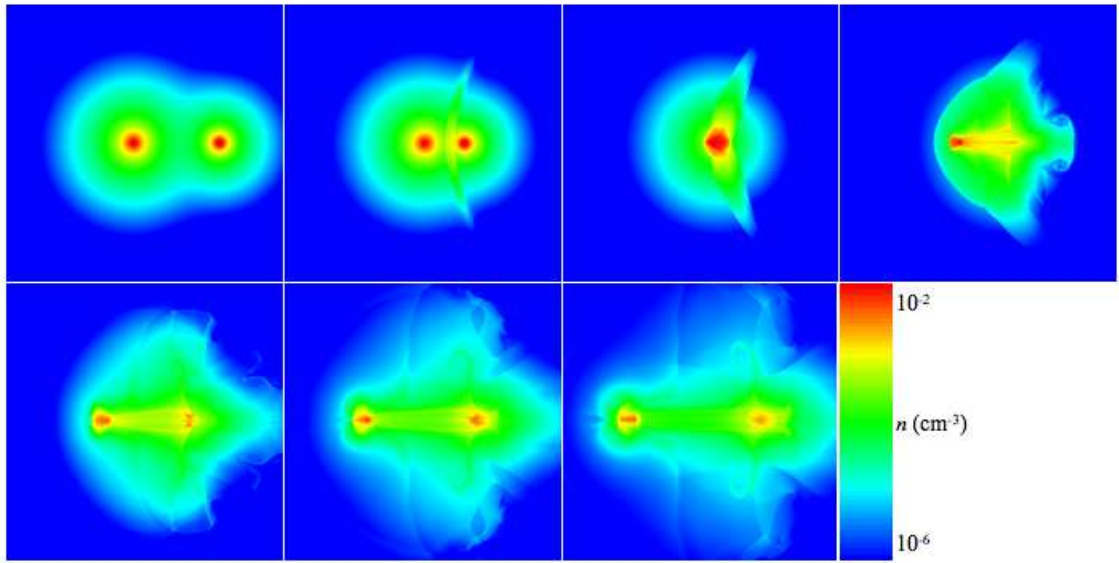}
\caption{Slices of gas density through the $z = 0$ coordinate plane for times $t$ = 0.0, 0.5, 1.0, 1.5, 2.0, 2.5, and 3.0 Gyr. The color scale is logarithmic. Each panel is 10 Mpc on a side.}
\end{figure*}


\begin{figure*}
\plotone{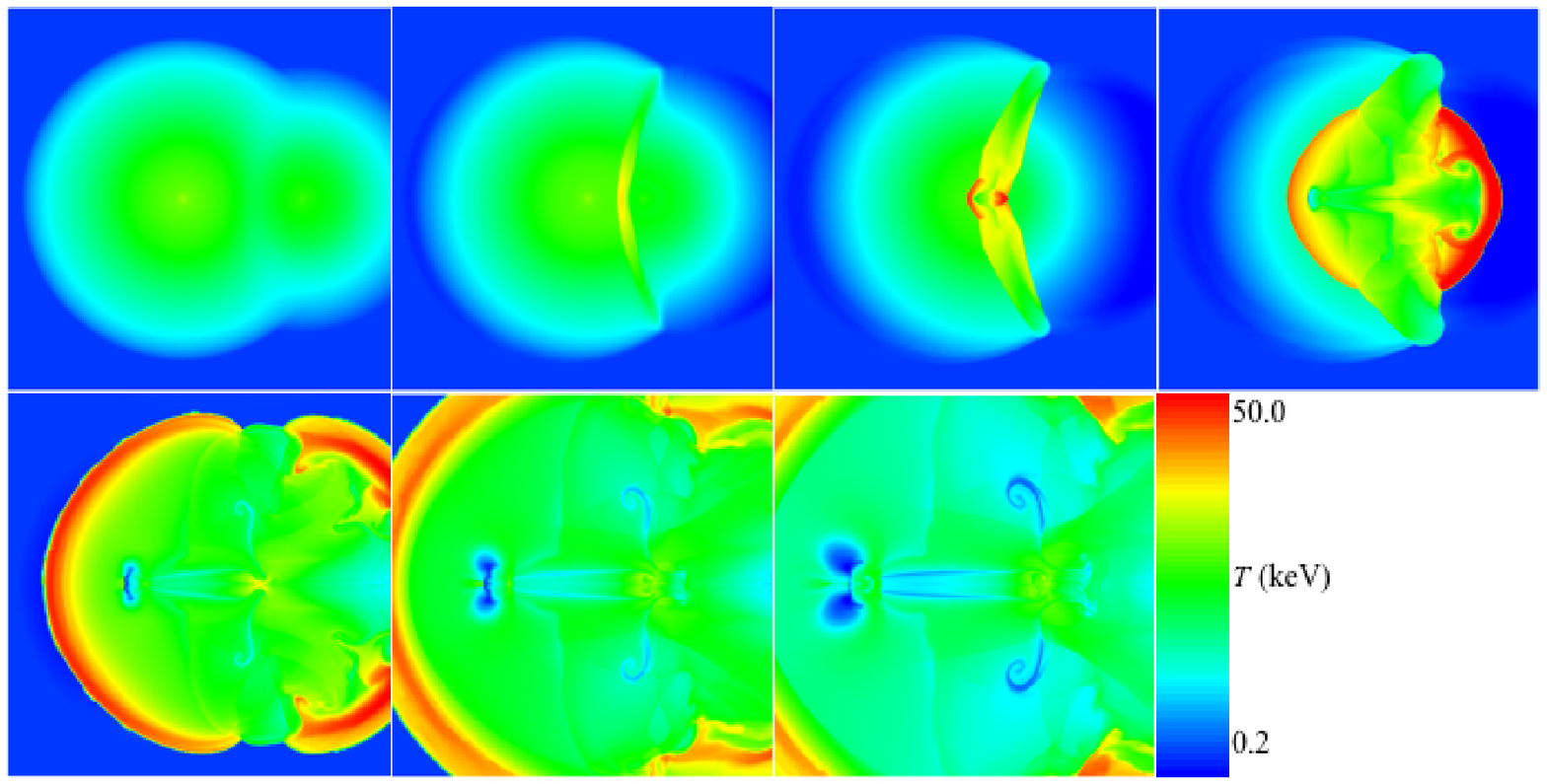}
\caption{Slices of gas temperature through the $z = 0$ coordinate plane for times $t$ = 0.0, 0.5, 1.0, 1.5, 2.0, 2.5, and 3.0 Gyr. The color scale is logarithmic. Each panel is 10 Mpc on a side.}
\end{figure*}


\begin{figure*}
\plotone{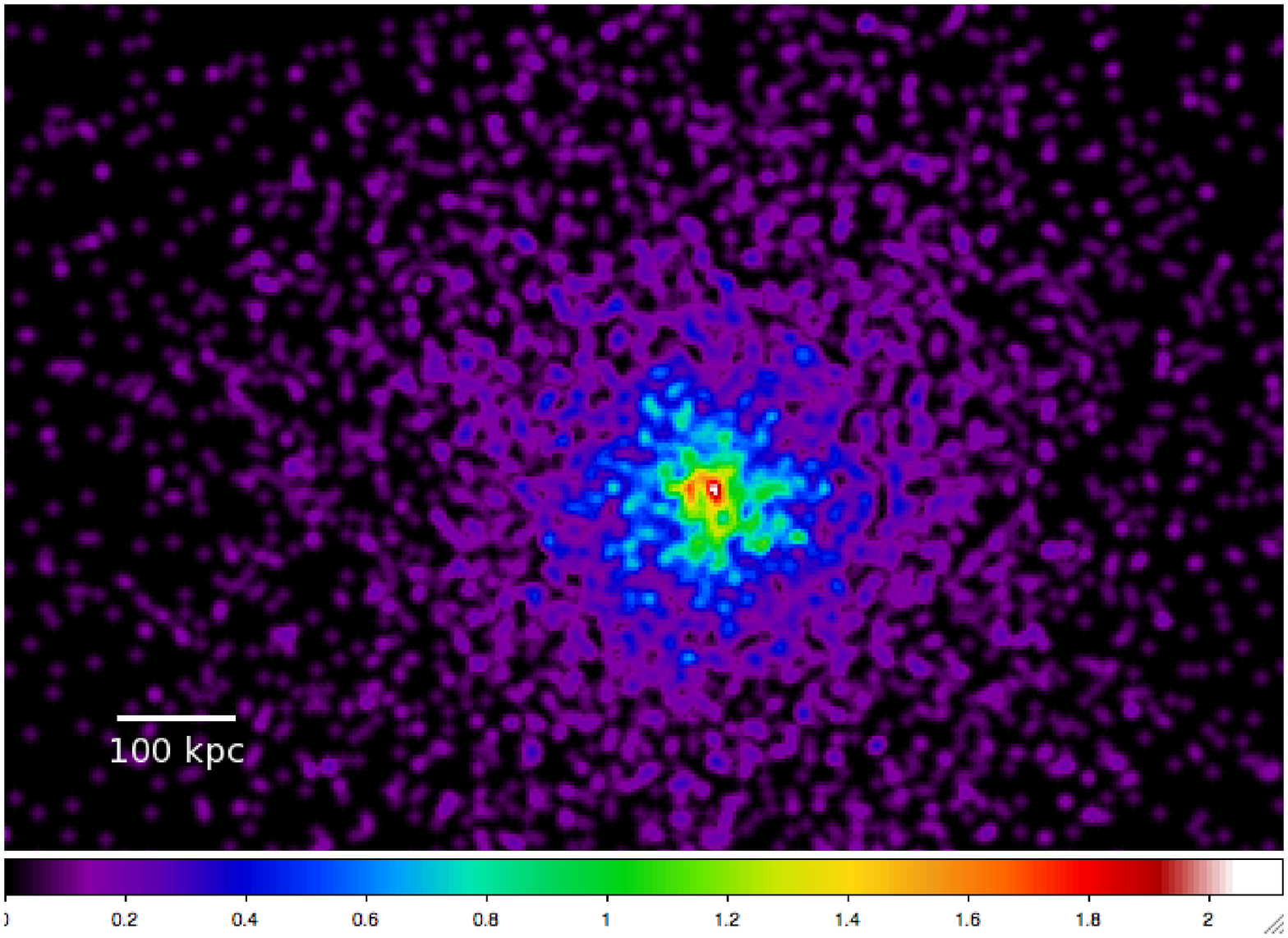}
\caption{Mock 40 ksec raw counts image of the simulation viewed along the line of centers of the clusters in the 0.5-5.0 keV band. The image has been smoothed using a Gaussian with a fixed kernel radius of 3 pixels. Color scale units are counts/pixel. The time is $t = 3.0$ Gyr, $\sim$2 Gyr after the collision.}
\end{figure*}

\section{Results}
 
\subsection{Qualitative Description}

Figures 1 and 2 show slices of density and temperature, respectively, through the $z = 0$ coordinate plane at different times in the simulation (the $x$ axis is the collision axis). From {\it t} = 0.0-0.9 Gyr the clusters approach each other and a shock front forms. At {\it t} = 1.0 Gyr the core of the smaller cluster collides with the core of the larger cluster, driving a shock wave forward into the larger core and displacing the larger core's gas from the collision axis into streams of cold gas. There is also a stream of dense, cold gas that is pulled directly between the two dark matter halos. Later on, the heads of these streams of gas, along with hotter gas from between the clusters, begin to fall into the potential of the larger halo and by {\it t} = 2.0 Gyr have fallen in completely. As for the gas core of the smaller cluster, it is penetrated shortly after the collision by a reverse shock, but this shock weakens as it traverses the core and so the latter remains cool, in agreement with similar investigations of high-velocity cluster mergers by \citet{tak05} and \citet{mil07}. In this process some of the gas of the smaller cluster is also ram-pressure stripped, and a contact discontinuity forms that is similar to the one observed in the 1E0657-56, and in the simulations of the latter \citep{spr07,mas08}. 

After penetrating the larger's gas core the smaller core is forced to lag behind its dark matter core by ram pressure until it encounters regions of lower density ({\it t} $>$ 1.2 Gyr). At this point the gas falls back toward the dark matter core but overshoots the potential and begins to slosh back and forth inside it. At late times the gas of the clusters is still settling into their respective dark matter potential wells, and the temperature structure of the gas is still complicated. Due to the high initial velocity of the collision, the clusters are in an unbound orbit and by the end of the simulation the separation between their respective dark matter halo centers is still increasing. This is in contrast to many previous works involving mergers \citep[e.g.,][]{ric01,poo06}, which assumed an initial velocity $v \approx v_{ff}$.


\begin{figure}
\plotone{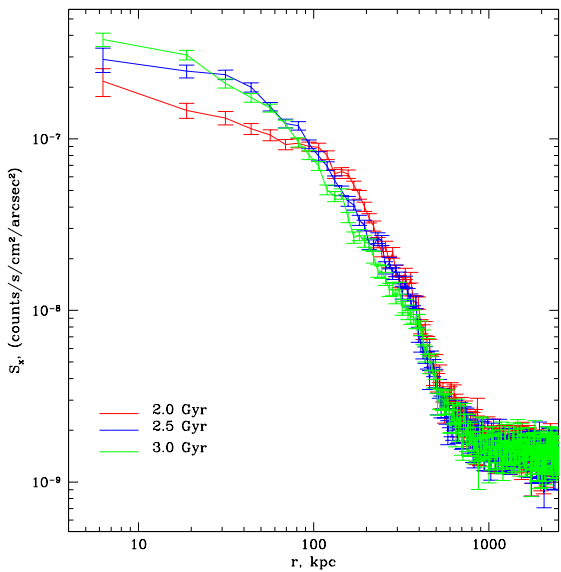}
\caption{Plots of surface brightness vs. projected radius for times $t =$ 2.0, 2.5, and 3.0 Gyr.}
\end{figure}


\begin{figure*}
\plottwo{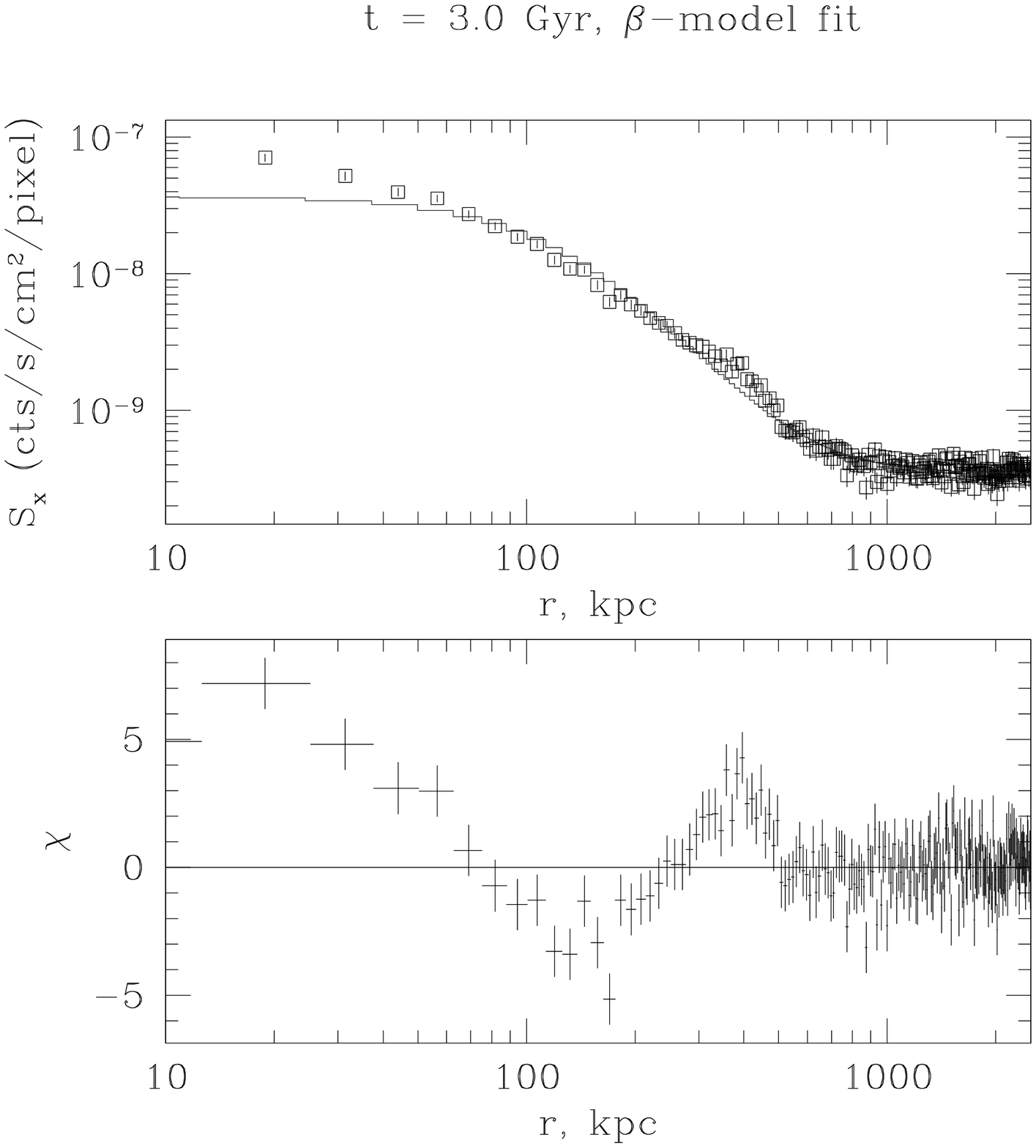}{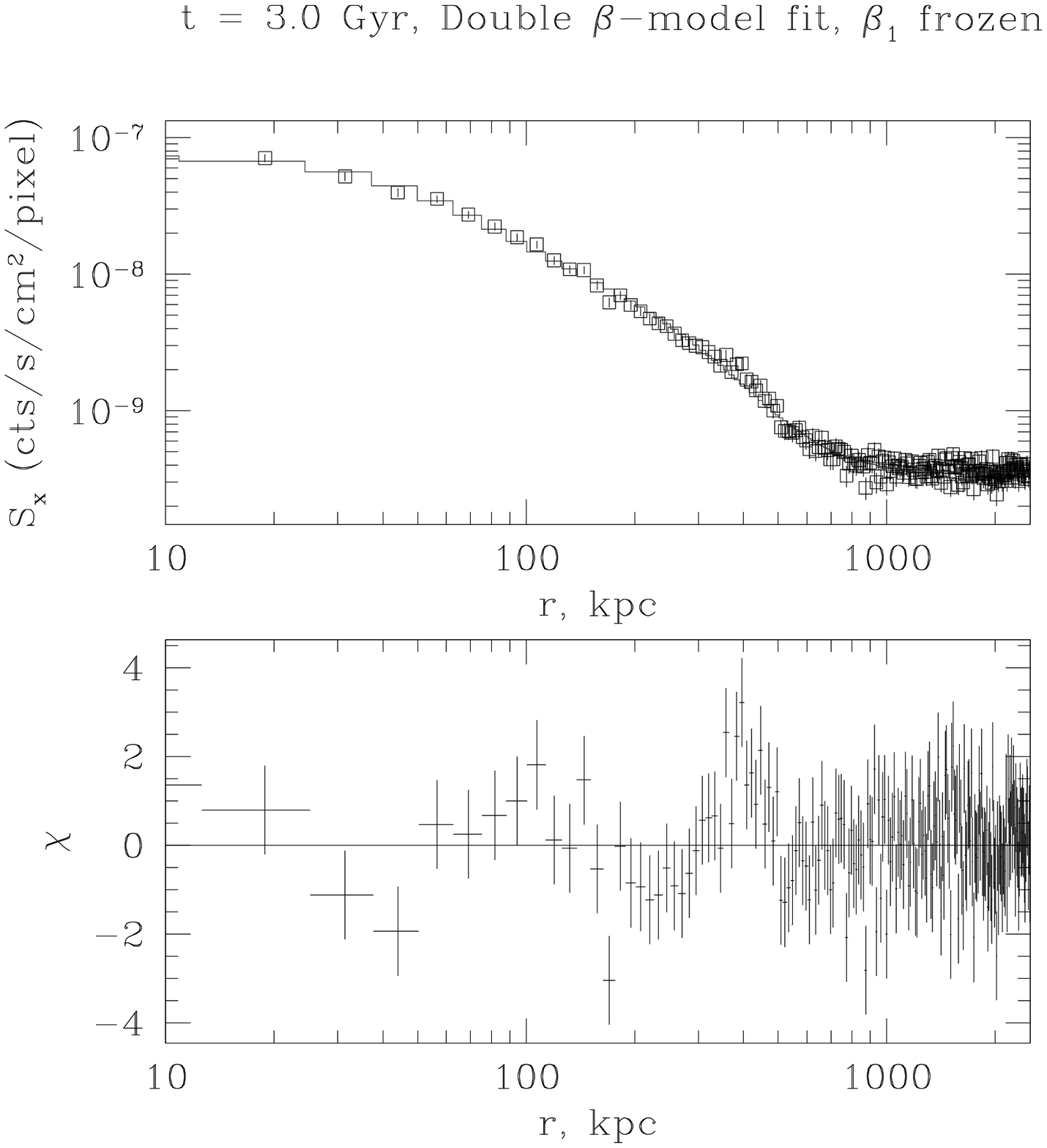}
\caption{Plots of single and double $\beta$-model fits for $t$ = 3.0 Gyr, $\sim$2 Gyr after the collision.}
\end{figure*}

\subsection{Mock X-ray Observations}

For the X-ray observations we view the system along the line of centers. Since we do not know the current state of the actual collision, we set each individual observation as if it were being observed at a redshift $z = 0.395$. At this redshift for the assumed cosmology 1'' = 6.39 kpc. Figure 3 shows an example of a smoothed mock X-ray image. For each observed time we extract a radial surface brightness profile and a radial temperature profile. 


\begin{figure}
\plotone{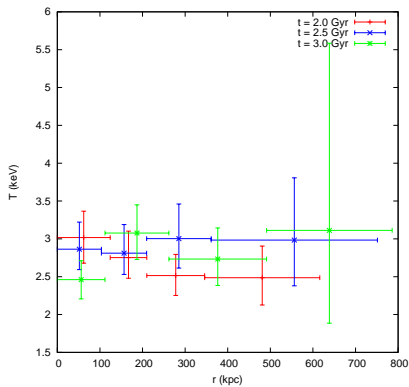}
\caption{Plot of temperature profile vs. projected radius for times $t =$ 2.0, 2.5, and 3.0 Gyr.}
\end{figure}

Cluster surface brightness profiles are extracted for the energy range 0.5-5.0 keV, within a circle centered on the surface brightness peak with a radius of 400'' ($\approx$ 2500 kpc). Figure 4 shows the surface brightness profiles for times $t =$ 2.0, 2.5, and 3.0 Gyr. The profiles are fitted with the $\beta$-model (Cavaliere \& Fusco-Femiano 1976):
\begin{equation}
S_X(r) = S_0\left[1+\left({r \over r_c}\right)^2\right]^{-3\beta+1/2}\ .
\end{equation}
We fit with a single $\beta$-model and a sum of two $\beta$-models. The $\chi^2$-statistic is used to fit the profiles. We follow \citet{ota04} and \citet{jee07} in fixing one of the $\beta$ parameters to unity, as there exists a strong degeneracy between this parameter and its corresponding core radius $r_{\rm c}$. We include the background constant as a free parameter in the single $\beta$-model fit, but freeze this value for the double $\beta$-model fit. The fitted profiles for the $t = 3.0$ Gyr case are shown in Figure 5. The values of the fitted parameters for epochs $t =$ 2.0, 2.5, and 3.0 Gyr are given in Tables 2 and 3. Errors are quoted at the 90$\%$ confidence level.

The projected temperature profiles for $t =$ 2.0, 2.5, and 3.0 Gyr are shown in Figure 6. The profiles were generated by extracting spectra from 4 annular regions in the 0.5-7.0 keV band centered on the peak of the surface brightness profile. Each is fitted with a MEKAL model, using the $\chi^2$-statistic with grouping to ensure at least 15 photons per bin. The radial extent of the temperature profiles is 125'' ($\approx$ 800 kpc), which is slightly greater than the radial range for which the temperature profile was extracted in \citet{ota04}. Single-temperature fits to the spectrum of the whole cluster in the 0.5-7.0 keV band for $t =$ 2.0, 2.5, and 3.0 Gyr are given in Table 4.


\begin{deluxetable*}{ccccc}
\tabletypesize{\scriptsize}
\tablecaption{Fitted Cluster Parameters for a Single $\beta$-model\label{tab2}}
\tablewidth{0pt}
\tablehead{
\colhead{$t$ ({\rm Gyr})} & 
\colhead{$r_{\rm c}$ ({\rm kpc})} & 
\colhead{$S_0$ (cts s$^{-1}$cm$^{-2}$arcsec$^{-2}$)} & 
\colhead{$\beta$} &
\colhead{$\chi^2$/d.o.f.}
}
\startdata
0.0 & $97.65_{-0.96}^{+0.18}$ & $(3.81_{-0.02}^{+0.02}) \times 10^{-6}$ & $0.798_{-0.003}^{+0.001}$ & 407.61/195 \\
2.0 & $172.04_{-1.73}^{+7.58}$ & $(1.42_{-0.03}^{+0.03}) \times 10^{-7}$ & $0.714_{-0.004}^{+0.017}$ & 305.14/195 \\ 
2.5 & $235.21_{-13.91}^{+2.24}$ & $(1.25_{-0.03}^{+0.03}) \times 10^{-7}$ & $0.964_{-0.031}^{+0.009}$ & 491.04/195 \\ 
3.0 & $150.24_{-12.50}^{+1.86}$ & $(1.43_{-0.03}^{+0.03}) \times 10^{-7}$ & $0.733_{-0.027}^{+0.005}$ & 423.58/195 \\ 
\enddata
\end{deluxetable*}


\begin{deluxetable*}{cccccccc}
\tabletypesize{\scriptsize}
\tablecaption{Fitted Cluster Parameters for a Double $\beta$-model\label{tab3}}
\tablewidth{0pt}
\tablehead{
\colhead{$t$ ({\rm Gyr})} & 
\colhead{$r_{\rm c_1}$ ({\rm kpc})} &  
\colhead{$S_{0,1}$ (cts s$^{-1}$cm$^{-2}$arcsec$^{-2}$)} &
\colhead{$\beta_1$} &
\colhead{$r_{\rm c_2}$ ({\rm kpc})} &  
\colhead{$S_{0,2}$ (cts s$^{-1}$cm$^{-2}$arcsec$^{-2}$)} & 
\colhead{$\beta_2$} &
\colhead{$\chi^2$/d.o.f.}
}
\startdata
0.0 & $89.24_{-0.32}^{+3.16}$ & $(3.73_{-0.04}^{+0.04}) \times 10^{-6}$ & 1.0 (F) & $204.65_{-4.00}^{+1.32}$ & $(7.03_{-0.07}^{+0.07}) \times 10^{-7}$ & $0.981_{-0.014}^{+0.003
}$ & 260.01/193 \\ 
2.0 & $39.91_{-17.45}^{+34.41}$ & $(8.18_{-3.39}^{+3.41}) \times 10^{-8}$ & 1.0 (F) & $270.00_{-4.78}^{+6.34}$ & $(1.11_{-0.02}^{+0.02}) \times 10^{-7}$ & $0.970_{-0.025}^{+0.0
10}$ & 245.83/193 \\ 
2.5 & $79.28_{-5.36}^{+9.04}$ & $(2.26_{-0.17}^{+0.17}) \times 10^{-7}$ & 1.0 (F) & $282.00_{-5.57}^{+11.97}$ & $(7.70_{-0.21}^{+0.21}) \times 10^{-8}$ & $0.970_{-0.021}^{+0.02
1}$ & 226.75/193 \\ 
3.0 & $82.12_{-3.94}^{+8.12}$ & $(2.51_{-0.16}^{+0.16}) \times 10^{-7}$ & 1.0 (F) & $316.24_{-4.89}^{+26.16}$ & $(4.94_{-0.16}^{+0.16}) \times 10^{-8}$ & $0.980_{-0.049}^{+0.01
5}$ & 205.86/193 \\
\enddata
\end{deluxetable*}


\begin{deluxetable}{ccc}
\tabletypesize{\scriptsize}
\tablecaption{Fitted Average Spectral Temperatures\label{tab4}}
\tablewidth{0pt}
\tablehead{
\colhead{$t$ ({\rm Gyr})} & \colhead{$T_{\rm spec}$ ({\rm keV})} & \colhead{$\chi^2$/d.o.f.}
}
\startdata
2.0 & $2.67_{-0.16}^{+0.17}$ & 140.62/177 \\ 
2.5 & $2.83_{-0.18}^{+0.23}$ & 147.98/174 \\
3.0 & $2.71_{-0.19}^{+0.20}$ & 128.07/167 \\
\enddata
\end{deluxetable}


\begin{figure}
\plotone{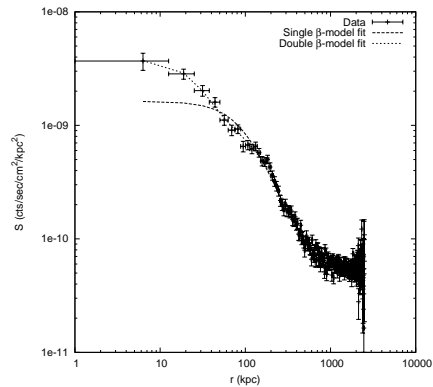}
\caption{Cl 0024+17 X-ray surface brightness profile and single and double $\beta$-model fits. Reproduced with permission from \citet{ota04}.}
\end{figure}

\section{Discussion}

\subsection{The Surface Brightness and Temperature Profiles}

The feature in the radial profile of Cl~0024+17 pointed out by \citet{ota04} and \citet{jee07} was that a double $\beta$-model fit is a better fit to the data, as a single $\beta$-model fit underestimates the central surface brightness. The surface brightness profile of Cl~0024+17 and the single and double $\beta$-model fits to the profile are shown in Figure 7. The difference between the fits is significant; there is a difference in the $\chi^2$-statistic of $\sim$60 for a difference of two degrees of freedom. 

In our mock cluster observations, we find that at later times, it is also possible to make a distinction between the two surface brightness components of the two clusters from the difference in the fitted $\chi^2$-statistic (see Tables 2 and 3). At the $t = 2.0$ Gyr epoch, $\Delta\chi^2 \approx$45 for a difference of only 2 degrees of freedom, and at later times is even larger ($\Delta\chi^2 \approx$200-260 for a difference of two degrees of freedom). We also note that we can use the $\Delta\chi^2$-test to distinguish between the two clusters in our initial conditions, which we also show in Tables 2 and 3. 


\begin{deluxetable*}{cccccc}
\tabletypesize{\scriptsize}
\tablecaption{Estimated and Exact Masses at R = 35"\label{tab5}}
\tablewidth{0pt}
\tablehead{
\colhead{$t$ ({\rm Gyr})} & \colhead{$M_{X,\beta} ({\rm M_\sun})$} & \colhead{$M_{X,2\beta} ({\rm M_\sun})$} & \colhead{$M_{\rm actual} ({\rm M_\sun})$} & \colhead{$M_{X,\beta}/M_{\rm actual}$} & \colhead{$M_{X,2\beta}/M_{\rm actual}$}}
\startdata
2.0 & $(6.03_{-0.36}^{+0.42}) \times 10^{13}$ & $(1.71_{-0.08}^{+0.08}) \times 10^{14}$ & $1.74 \times 10^{14}$ & 0.35 & 0.98 \\ 
2.5 & $(7.50_{-0.58}^{+0.61}) \times 10^{13}$ & $(1.74_{-0.08}^{+0.10}) \times 10^{14}$ & $1.65 \times 10^{14}$ & 0.45 & 1.05 \\ 
3.0 & $(6.57_{-0.54}^{+0.49}) \times 10^{13}$ & $(1.62_{-0.08}^{+0.10}) \times 10^{14}$ & $1.55 \times 10^{14}$ & 0.42 & 1.05 \\ 
\enddata
\end{deluxetable*}


\begin{figure*}
\plotone{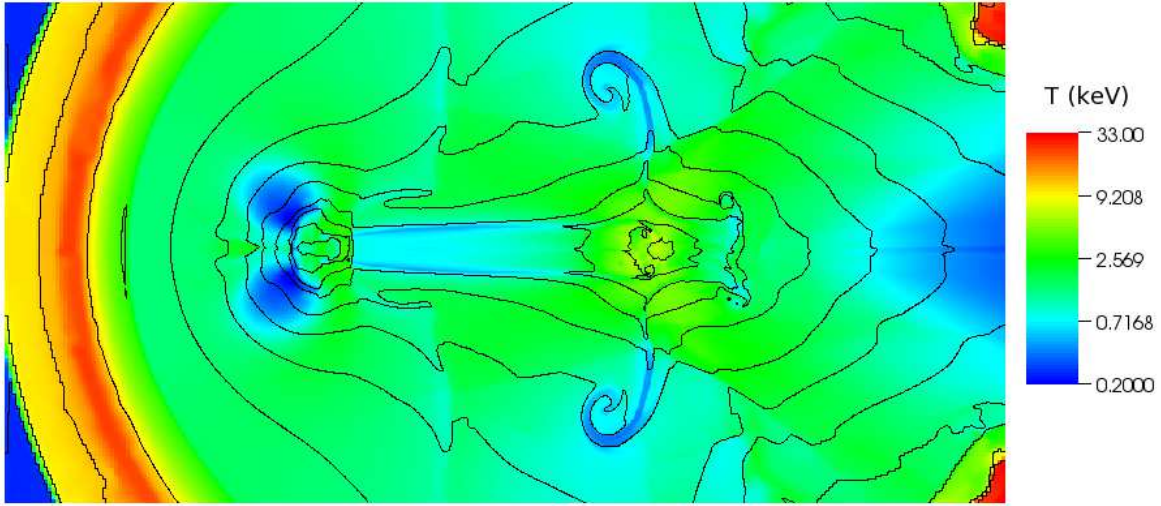}
\caption{Slice through the $z = 0$ coordinate plane of the simulation of temperature at $t = 3.0$ Gyr, $\sim$2 Gyr after the collision. The colorscale is logarithmic. Density contours are logarithmically spaced by 2.5$\times$, between $n = 10^{-6}-10^{-2} {\rm cm}^{-3}$. The size of the figure is 14 Mpc by 6 Mpc.}
\end{figure*}


\begin{figure}
\plotone{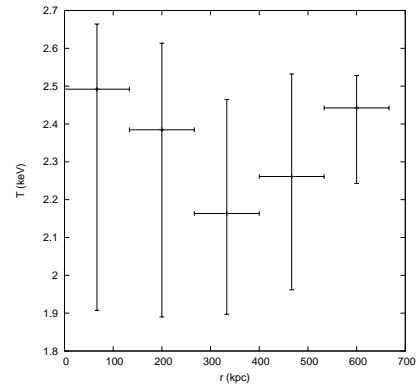}
\caption{Radial profile of mass-weighted temperature at $t = 3.0$ Gyr, $\sim$2 Gyr after the collision.}
\end{figure}


\begin{figure}
\plotone{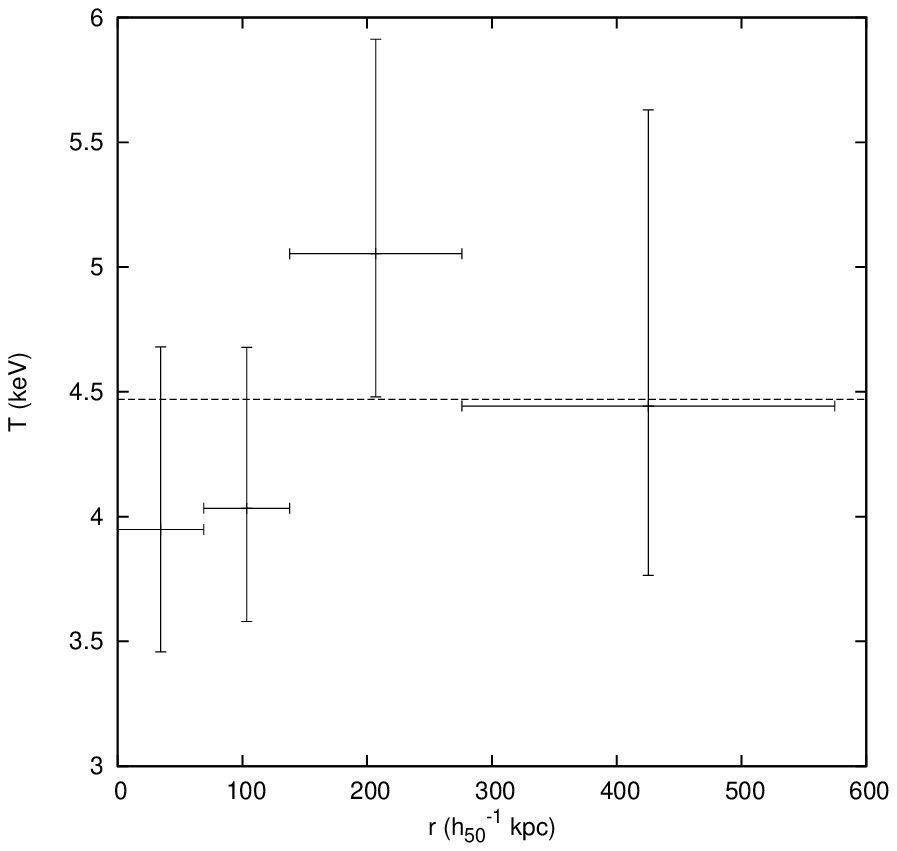}
\caption{Cl~0024+17 Temperature profile. Average profile is shown as a dashed line at $T = 4.47$ keV. Reproduced with permission from \citet{ota04}.}
\end{figure}

In Cl~0024+17, the fitted spectral temperature of $T = 4.47$ keV seems low for a system undergoing a collision or merger. In our mock observations we note a similar phenomenon; the fitted temperatures in the mock observations are significantly lower than what might be expected given the temperatures observed in the simulation. In our simulated system, the hottest cluster gas is in the larger cluster and is at a temperature of $\sim$ 5-6 keV at later times. However, the measured temperatures in the mock X-ray observations barely reach $ \sim$ 3~keV. A close look at the temperature distribution in the simulation itself (Figure 8) reveals that the highest-temperature gas is confined to the central region of the larger cluster, and the denser gas of the smaller cluster is colder, around $3$ keV. In projection along the line of sight, the lower-temperature, denser gas ``washes out'' the higher-temperature, less dense gas, and the complicated, two-component temperature structure of the clusters is lost because of the strong density dependence of the X-ray emission ($\propto \rho^2$). Figure 9 shows the profile of the mass-weighted temperature over the same range as Figure 6 for $t = 3.0$ Gyr. The lack of temperatures higher than $\sim 3$ keV demonstrates that the denser, colder gas provides the dominant contribution to the X-ray emission.

The single-temperature fits are in general agreement with our projected temperature profiles. Attempting to fit the cluster temperature using a two-temperature model resulted in no significant decrease in the $\chi^2$-statistic from the single-temperature fits, so on the basis of this analysis we cannot distinguish the two different average temperatures of the two clusters.

At late times there is still some moderate variation in the projected temperature profile with time, indicative of the continuing evolution of the system, though the average temperature of the cluster is roughly constant with time. In \citet{ota04}, the observed isothermal temperature profile of Cl~0024+17 (reproduced in Figure 10) was taken as evidence that the merger was not recent (i.e., it was more than a few Gyr ago). Our results show that if viewed along the line of sight $\sim$2 Gyr after the merger, the temperature structure of the system can appear more regular in projection than it actually is (e.g., the structure seen in Figure 2). If the results of \citet{jee07} regarding a ringlike dark matter structure indicate the clusters are being viewed along the line of sight shortly after the merger, our results show that the system can still appear relatively relaxed. This agrees with previous results of mock X-ray observations of simulations of galaxy cluster mergers (e.g., Ricker and Sarazin 2001, Poole et al. 2006).

Finally, our observed spectral temperature from our simulations of $T \sim 2.6-2.8$ keV is somewhat lower than the $T \sim 4.47$ keV for Cl~0024+17 measured by \citet{ota04} or the $T \sim 4.25$ keV measured by \citet{jee07}. If the collision scenario for Cl~0024+17 is correct, this indicates that the initial conditions for our simulation are not identical to the pre-merger conditions for the real cluster. A higher impact velocity or a change in the masses of the clusters could account for the temperature difference.

\subsection{The Reliability of Hydrostatic Mass Estimates}

Though the temperature and surface brightness profiles are undergoing considerable evolution even at late times, it is still relevant to ask how accurate a mass estimate would be under the assumption of hydrostatic equilibrium. As previously noted, \citet{ota04} and \citet{jee07} made estimates of the mass of Cl~0024+17 using single and double $\beta$-model fits to the surface brightness profile together with an isothermal temperature profile. Although our temperature profile is not strictly isothermal, existing methods for temperature deprojection assume spherical symmetry, so we do not attempt such a deprojection here. If we assume a cluster temperature $T_{\rm spec}$ given by the spectral temperature fit to the entire cluster, we can arrive at a rough estimate for the mass of the system under the assumption of hydrostatic equilibrium.

For a $\beta$-model fit to the cluster gas, the projected cluster mass within a cylindrical volume for a given temperature, $\beta$, and core radius $r_{\rm c}$ can be estimated by \citep{ota98,jee05}
\begin{equation}
M_{X,\beta}(R) = 1.78 \times 10^{14}\beta\left({T \over {\rm keV}}\right)\left({R \over {\rm Mpc}}\right){R/r_{\rm c} \over \sqrt{1+(R/r_{\rm c})^2}}{\rm M}_\sun\ ,
\end{equation} 
where $R$ is the projected radius, $T$ is the gas temperature and $\beta$ is the $\beta$-model index. We compute the estimated mass within the arc radius of the lensed galaxy in Cl~0024+17, $r_{\rm arc} = 35"/223.65$ kpc. For the double $\beta$-model fits we assume the same temperature for both cluster components, and we add the mass contributions from the two fits together. Table 5 shows the estimated masses from the single and double $\beta$-model fits for times $t =$ 2.0, 2.5, and 3.0 Gyr. Errors are quoted at the 90$\%$ confidence level. It is clear from the table that assuming a single-cluster model for the system will underestimate the projected mass by a factor $\sim$2-3, whereas assuming a double-cluster model will estimate the projected mass to within $\sim$10$\%$. This is in agreement with the results of \citet{jee07}. 

\subsection{The Effect of A Slightly Off-Axis Collision}

In our simulated head-on collision, the gas in the clusters is significantly disrupted and by the end of the simulation is still settling into the potential wells created by the dark matter. If instead of colliding head-on the encounter were allowed to be slightly off-axis, this would result in less disruption of the cluster gas cores. Presumably the gas would relax more quickly and an even more accurate mass determination could be made under the assumption of hydrostatic equilibrium. In addition, in a slightly off-axis collision viewed along the line of sight the two cluster components would be more clearly distinguished in the X-ray emission. Two-component fits to both the X-ray surface brightness and the temperature of the different components would enable a more accurate mass determination. This is a large parameter space of simulations that will be explored in future investigations, but is beyond the scope of this paper.


\begin{deluxetable*}{ccccccccc}
\tabletypesize{\scriptsize}
\tablecaption{Fitted parameters for Redshift Histograms with Varying $\sigma_z$\label{tab6}}
\tablewidth{0pt}
\tablehead{
\colhead{$\sigma_z$} & 
\colhead{$\mu_0$} & 
\colhead{$\sigma_0$} & 
\colhead{${\chi^2/{\rm d.o.f.}}_{single}$} &
\colhead{$\mu_1$} & 
\colhead{$\sigma_1$} & 
\colhead{$\mu_2$} & 
\colhead{$\sigma_2$} & 
\colhead{${\chi^2/{\rm d.o.f.}}_{double}$}
}
\startdata
0.005 & 0.9975 & 0.0040 & 12.6/10 & 1.002 & 0.0040 & 0.9959 & 
0.0052 & 5.88/7 \\
0.01  & 0.9958 & 0.0086 & 31.68/9 & 0.9998 & 0.0086 & 0.9889 & 
0.0089 & 14.7/6 \\
0.02  & 0.9936 & 0.0175 & 8.64/9 & 0.9947 & 0.0175 & 0.9742 & 
0.0126 & 9.84/6 \\
0.08  & 0.9521 & 0.0634 & 6.65/7 & 0.9733 & 0.0634 & 0.9260 & 
0.0794 & 3.56/4 \\
\enddata
\end{deluxetable*}


\begin{figure*}
\plottwo{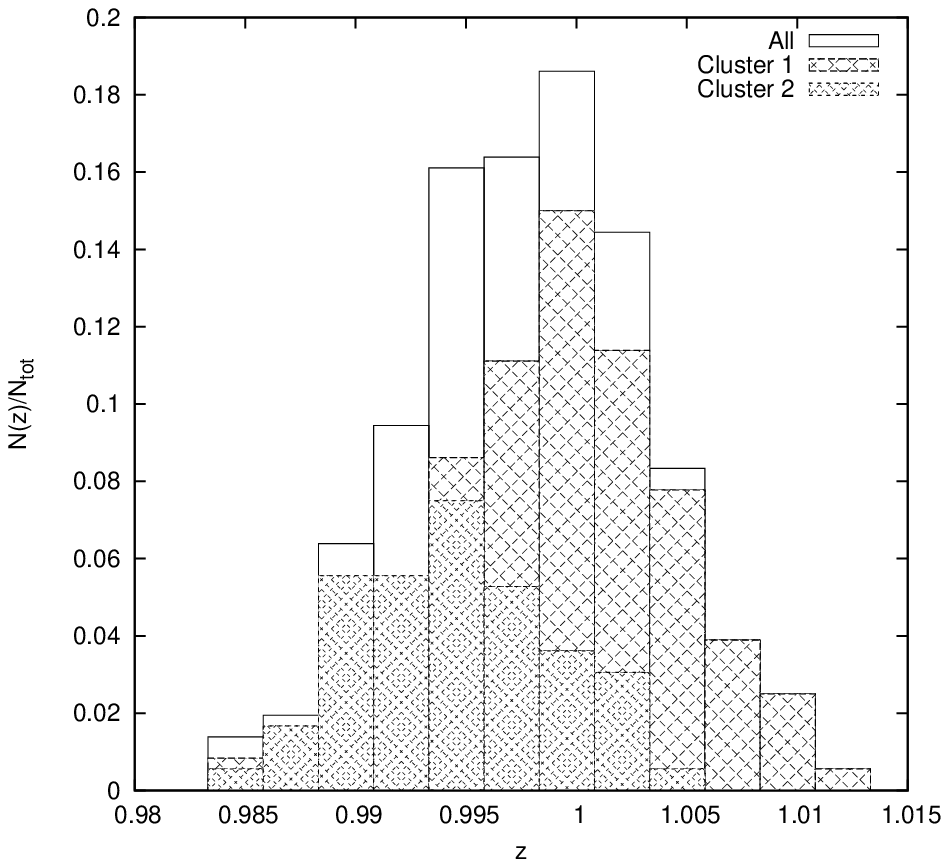}{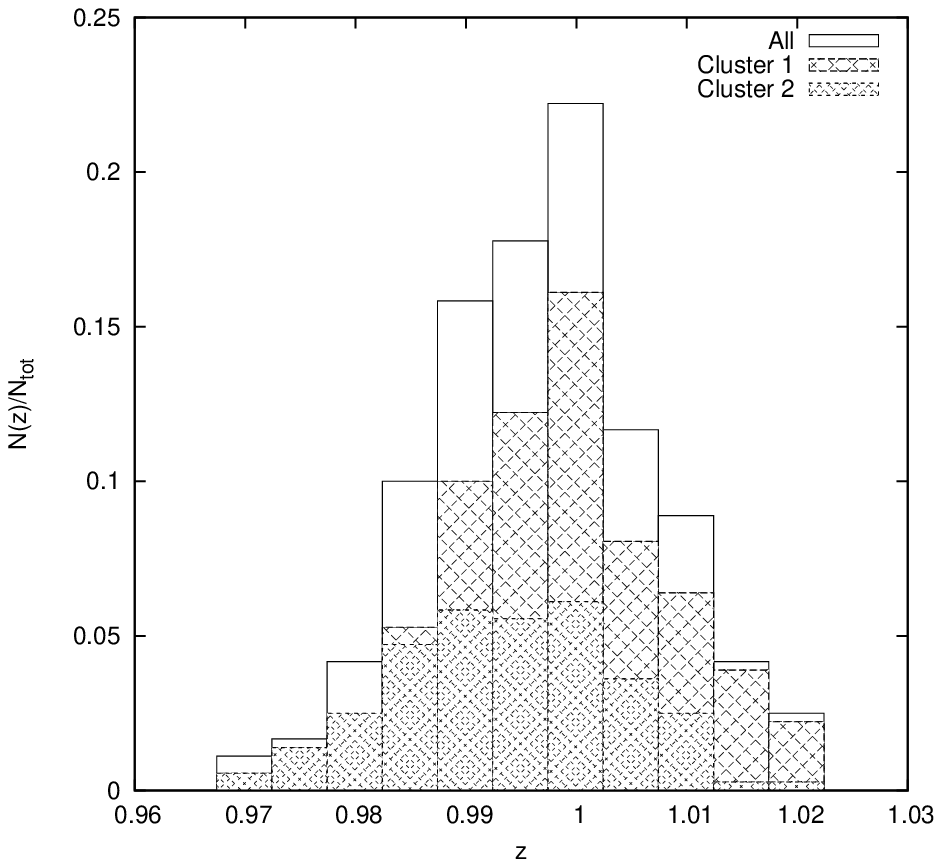}
\caption{Number of redshifts per bin vs. redshift, ${\sigma_z} = 0.005$ and ${\sigma_z} = 0.01$.}
\end{figure*}


\begin{figure*}
\plottwo{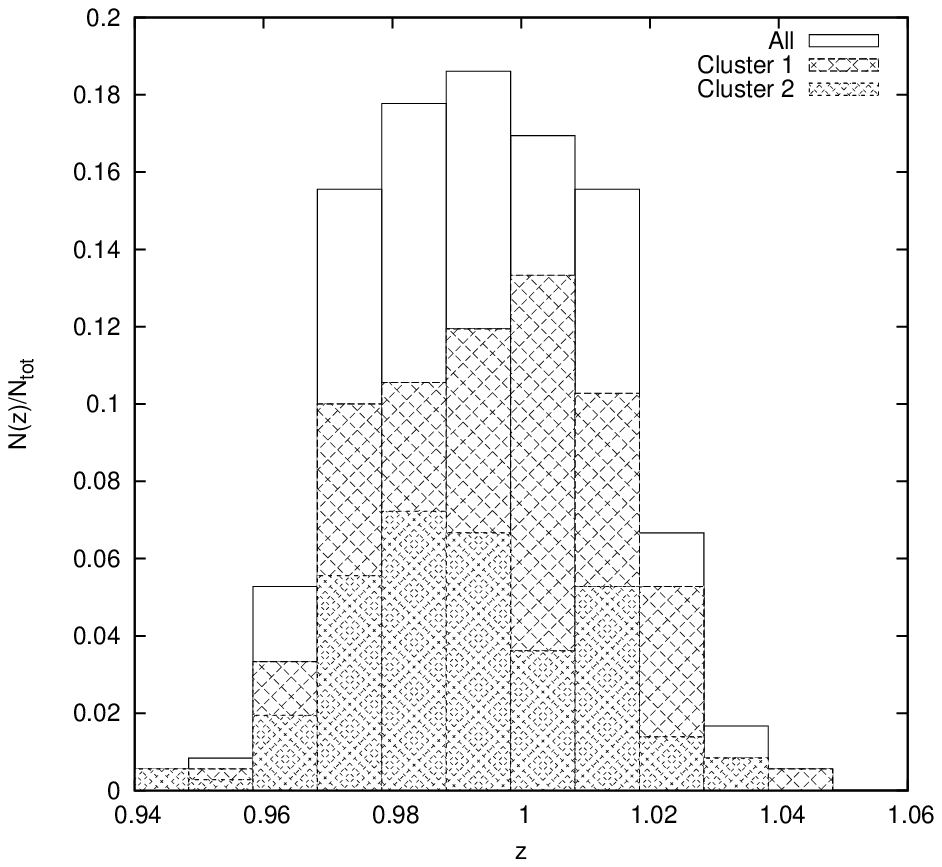}{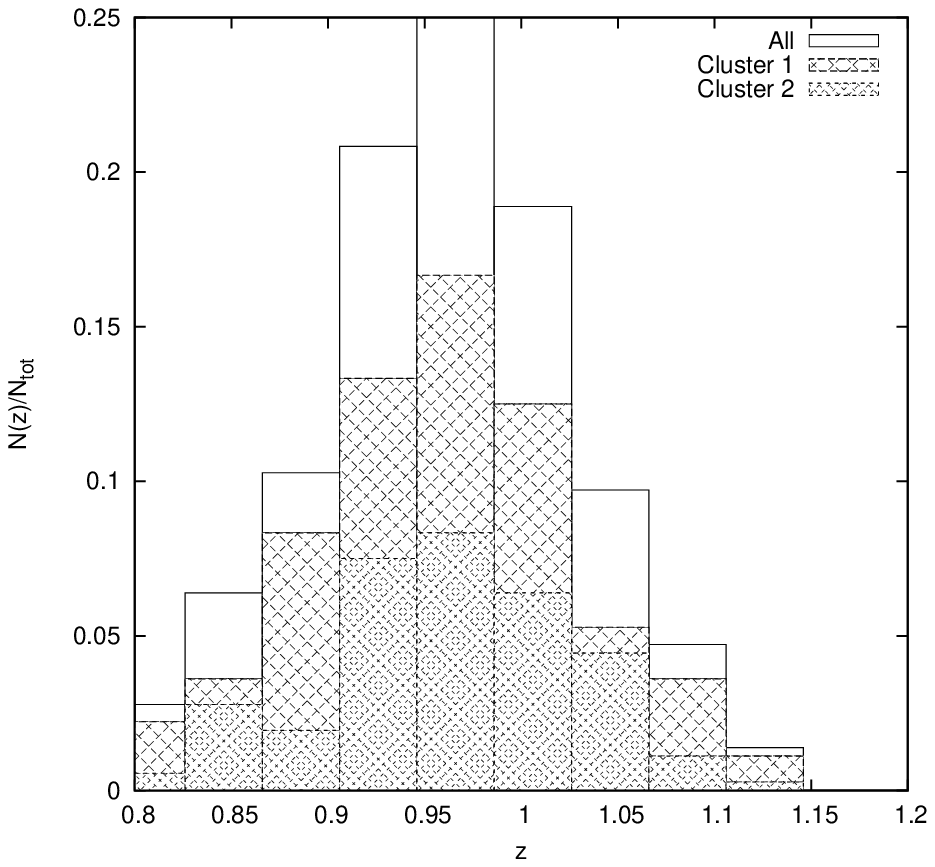}
\caption{Number of redshifts per bin vs. redshift, ${\sigma_z} = 0.02$ and ${\sigma_z} = 0.08$.}
\end{figure*}

\subsection{Implications for X-ray Surveys}

These results have important implications for X-ray surveys. Neighboring galaxy clusters viewed in superposition, whether undergoing mergers or not, may bias temperature measurements based on spectral fitting. This can have an effect particularly on X-ray-based scaling relations. For example, previous simulated observations of merging clusters have demonstrated that for mergers occurring along the line of sight, the clusters appear to have higher temperatures for their mass \citep{ric01, poo07}. Our investigation shows that clusters viewed in superposition can make a cluster appear colder than it actually is if the colder cluster's gas is significantly denser. In addition, discrepancies between hydrostatic and lensing mass estimates may in some cases be attributed to the existence of multiple components viewed along the line of sight. as is apparent in the case of Cl~0024+17. Detailed analysis of X-ray observations would be required to discern such superpositions.

One possible way to correct for and identify such superposition effects is by using the optical redshifts of the cluster galaxies to identify separate cluster components, as \citet{czo01,czo02} did in the case of Cl~0024+17. To reliably identify such separate cluster components requires accurate redshifts. Because spectroscopic redshifts are only feasible for the nearest or brightest cluster galaxies, cluster surveys must rely on photometric redshift estimators. Large optical surveys such as zCOSMOS \citep{lil07}, Combo-17 \citep{wol03}, and the upcoming Dark Energy Survey \citep{des05} have photometric redshift errors $\sigma_z \sim (0.02-0.05)(1+z)$.


\begin{deluxetable*}{ccccccc}
\tablecolumns{7}
\tabletypesize{\scriptsize}
\tablecaption{$\Delta\chi^2$ for Different Exposure Times and Redshifts\label{tab7}}
\tablewidth{0pt}
\tablehead{
\colhead{} & \colhead{} & \multicolumn{2}{c}{$t$ = 3.0~Gyr} &
\colhead{} & \multicolumn{2}{c}{$t$ = 2.0~Gyr} \\
\cline{3-4} \cline{6-7} \\
\colhead{$t_{\rm exp}$ (ks)} & \colhead{} & \colhead{$z = 0.395$} & \colhead{$z =
  1.0$} & \colhead{} & \colhead{$z = 0.395$} & \colhead{$z =
  1.0$}
}
\startdata
10 & & 177.69/96 - 113.21/94 & 99.20/96 - 86.05/94 & & 96.41/96 - 94.40/94 & 115.42/96 - 111.74/94 \\
40 & & 425.31/195 - 203.44/193 & 242.51/195 - 173.29/193 & &
305.14/195 - 245.83/193 & 217.34/195 - 209.53/193 \\
240 & & 1497.05/195 - 439.33/193 & 770.37/195 - 287.46/193 & &
428.73/195 - 245.83/193 & 280.26/195 - 228.22/193 \\
\enddata
\end{deluxetable*}

If the only redshift determination for Cl~0024+17 were via photometry, would the separation between the two components still be discernible? To answer this question we constructed a mock ``galaxy'' catalog from our simulation. We chose dark matter particles from the two clusters as proxies for the cluster galaxies (120 from the smaller cluster, 240 from the larger) and assumed a value for $\sigma_z$.  We constructed redshift histograms for the resulting galaxy distributions and fitted them to both a Gaussian distribution and a sum of two Gaussian distributions:
\begin{equation}
f_{single} = A_0e^{-(z-{\mu_0})^2/2{\sigma_0^2}}
\end{equation}
\begin{equation}
f_{double} = A_1e^{-(z-{\mu_1})^2/2{\sigma_1^2}} + A_2e^{-(z-{\mu_2})^2/2{\sigma_2}}\ .
\end{equation}
The fitted parameters for these distributions for varying $\sigma_z$, given a choice of mock galaxies from our simulation, are shown in Table 6. For low values of $\sigma_z$, the mean values of the two redshift distributions are statistically distinguishable, but for higher values they are not. The histograms shown in Figures 11 and 12 demonstrate that for $\sigma_z \sim 0.005-0.01$ the skewness of the distribution of galaxies is evident, but when the redshift error increases to $\sigma_z \simgt 0.02$, the resulting galaxy distribution is indistinguishable from that of a single cluster.

At a redshift of $z \approx 0.4$ (Cl~0024+17), the expected photometric redshift errors from optical surveys would be too large at this time to distinguish between the two cluster components ($\sigma_z \sim 0.03-0.07$), and at a redshift of $z \approx 1$ it would be worse ($\sigma_z \sim 0.04-0.1$). Identification of line-of-sight mergers and collisions from redshift distributions likely will be restricted to clusters for which we have spectroscopic redshifts available.

Our X-ray fitting results might be taken to suggest that X-ray surface brightness profiles could be used in place of galaxy redshift distributions to detect clusters seen in superposition. To test this hypothesis, we placed our simulated clusters at the epoch $t$ = 3.0 Gyr ($\sim$2 Gyr after the collision, where we can clearly distinguish both cluster components) at two redshifts, the current redshift of $z = 0.395$ and a redshift of $z = 1.0$. The same analysis of the surface brightness profiles as above was performed for exposure times of 10 ks, 40 ks, and 240 ks, the former being a more typical exposure time for a given cluster in an X-ray survey. We focus here on the difference in the $\chi^2$ between the single and double $\beta$-model fits. The resulting $\Delta\chi^2$ for the fitted surface brightness models corresponding to the different exposure times for the different redshifts are shown in Table 7. At $z = 0.395$, it is possible to use the $\Delta\chi^2$ between the two models to distinguish between the two clusters for all exposure times. At $z = 1.0$, for an exposure time of 10 ks, the $\Delta\chi^2$ is $\approx 14$ for a change of 2 degrees of freedom, and for 40 ks $\Delta\chi^2 \approx 70$. Therefore, we find that at exposure times more typical of X-ray surveys, even at high redshift we can distinguish between the two cluster components, though with more difficulty. This is potentially a promising avenue for identifying merging clusters and clusters in projection. However, the ability to identify merging clusters requires that the two clusters have surface brightness profiles that are distinct enough that the likelihood-ratio test can distinguish between them.  For example, earlier on in the simulation, at the epoch $t$ =  2.0~Gyr ($\approx$ 1.0~Gyr after the collision, where we cannot distinguish between the two clusters as clearly), we find that the test can distinguish between the two components for our fiducial exposure time and redshift, but not at lower exposures and higher redshifts (Table 7). Thus, determining exactly the physical circumstances under which the likelihood ratio test is the most powerful way to identify merging clusters and clusters in projection requires further study.

\section{Conclusions}

The galaxy cluster Cl~0024+17 is thought to be a colliding system of galaxy clusters that is viewed along the collision axis. We have performed a simulation of a high-velocity head-on collision of galaxy clusters and created mock X-ray observations, viewing the clusters from the same direction. Previous investigations of this system had focused on the collisionless dynamics of the galaxies and the dark matter; we also include gas in our simulation. The mock X-ray observations of our simulation indicate that the X-ray emitting gas is still undergoing moderate evolution at times $\sim$1-2 Gyr after the collision. However, much of the complicated structure in density and temperature is obscured due to projection effects. The X-ray surface brightness profile is better fit by a superposition of $\beta$-models than a single $\beta$-model at times $t =$ 1-2 Gyr after the collision, corresponding to the observed situation in Cl~0024+17. The cluster gas in the center appears colder than it really is due to the projection of denser, colder gas along the line of sight of the hotter gas. The temperature profile exhibits only moderate evolution with time, though the temperature distribution of the clusters in the simulation has significant structure. Though a mass estimate of the cluster assuming one cluster component in hydrostatic equilbrium underestimates the mass by a factor of $\sim$2-3, a mass estimate based on the assumption of two galaxy clusters in superposition comes close ($\sim$10$\%$) to the actual projected mass of the system within the arc radius $r_{\rm arc}$, in agreement with the mass measurements made of the cluster Cl~0024+17 under the same assumptions. These results may be valuable when looking at clusters discovered in X-ray surveys. With current photometric redshift errors it will not likely be possible to distinguish such line-of-sight collisions from single clusters discovered in X-ray surveys at high redshift by identifying separate galaxy concentrations. However, our results show that it may be possible to distinguish merging clusters and clusters in projection from single clusters by testing multiple model fits against single model fits, though this will become more difficult at higher redshifts.

\acknowledgments
This work is supported at the University of Chicago by the U.S Department of Energy (DOE) under Contract B523820 to the ASC Alliance Center for Astrophysical Nuclear Flashes. Calculations were performed using the computational resources of Lawrence Livermore National Laboratory. JAZ is grateful to Andrey Kravtsov, Carlo Graziani, John Davis, Joe Mohr, and Maxim Markevitch for useful discussions and advice. PMR and HYY acknowledge support under a Presidential Early Career Award from the U.S. Department of Energy, Lawrence Livermore National Laboratory (contract B532720) and by NASA grant NNX06AG57G. JAZ is supported by the Department of Energy Computational Science Graduate Fellowship, which is provided under grant number DE-FG02-97ER25308.

\clearpage

\appendix

\section{Mock Observation Generation and Verification Study}

Our mock X-ray observation generation procedure consists of two steps: we first use the FLASH input of density and temperature to create a flux map in sky coordinates for a range of energies. In doing this we follow closely the procedure outlined in \citet{gar04}. Secondly, we use this flux map as a ``user source'' in MARX to generate the photon energies, positions, and times for our simulated observation.

The FLASH AMR grid is regridded to a uniform grid at the highest resolution in the simulation. We then choose a direction along which the cluster is observed and the physical quantities are projected. For each cell in our dataset the emissivity of the plasma is given by
\begin{equation}
\epsilon = n_en_H\Lambda(T,Z)
\end{equation}
where $n_e$ and $n_H$ are the electron and hydrogen densities, respectively, and $\Lambda(T,Z)$ is the power coefficient which depends on temperature and metallicity, which are assumed constant over one FLASH cell size. This coefficient is calculated using the MEKAL model. Using this relation, the photon luminosity (photons s$^{-1}$ cm$^{-2}$ keV$^{-1}$) at a given energy $h\nu$ is given by
\begin{equation}
L^\gamma_\nu = \int_V{\epsilon^\gamma_\nu}dV' = {\Lambda^\gamma_\nu}EM
\end{equation}
where the quantity $EM \equiv \int_V{n_en_H}dV'$ is the emission measure. The measured flux of photons at a specific energy at a redshift $z$ is given by
\begin{equation}
F^\gamma_\nu = \frac{(1+z)^2L^\gamma_{\nu(1+z)}}{4{\pi}D^2_L}
\end{equation}
where $D_L$ is the luminosity distance for the given redshift and cosmology. 

This flux map is then used to generate photons in MARX using the ``user source'' implementation. The map is taken as an input and used as a distribution function for the photons to determine their position in sky coordinates, energy, and time of detection. In the cases considered here, the FLASH zone size is larger (typically by a factor of $\sim$4) than the size corresponding to a Chandra pixel. Within this FLASH zone size the photon positions are uniformly distributed. After the events file is generated, we re-bin the image by this factor to perform our spatial analysis.

For the spatial analysis, we construct an exposure map for the ACIS-S
chips with the input spectrum of the cluster as a set of weights, and
when performing the fit we convolve the model with a 1-D PSF. For the
spectral analysis, we construct weighted RMFs and ARFs (Redistribution
Matrix Files and Auxiliary Response Files), and use them to fit a background-subtracted spectrum. Our spectra are grouped so that each bin has at least 15 counts.
 
In order to ensure the accuracy of our MARX simulations, we have performed a few simple verification tests. We have constructed two models: one, a gas sphere with a constant temperature and a $\beta$-model density profile, and the other, two gas spheres, each with different temperatures and $\beta$-model parameters. We then subjected them to the same analysis as our simulated clusters in FLASH. In the single-cluster case, we verify that the surface brightness fits we recover the core radius $r_{\rm c}$, $\beta$-parameter, and central surface brightness $S_0$. Also, we verify in the spectral fit that we recover the model temperature for the fit to the whole cluster, the normalization of the spectrum, and the isothermal temperature profile. For the two-cluster case, we verify that the two cluster components can be distinguished and that the corresponding $\beta$-model parameters are recovered (as in the simulated clusters, we find that we cannot distinguish a single spectral temperature model from a sum of spectral temperature models). For these tests we consider the model verified if we can recover the model parameters within the 1$\sigma$ errors.

For the spatial fitting, we find that as we increase the exposure time of the observation, systematic effects become more important. At an exposure time of $t_{\rm exp}$ = 240 ks we recover the fitted parameters for both the single and double $\beta$-model cases. This corresponds to the highest exposure time for our simulated clusters. These results are shown in Tables 8 and 9. However, as we go to higher exposure times, we find that we cannot recover the input models. Table 11 shows the number of parameters which are not recovered and the corresponding reduced $\chi^2$-statistic for increasing exposure time. For an exposure time of 1 Ms, we can verify the single $\beta$-model but cannot verify the double $\beta$-model, and for an exposure time of 4 Ms we cannot verify either. 

Possible systematic effects may include the modeling of the PSF (which is energy and spatially dependent), ACIS read-out errors and the precise dependence of the exposure map on the input spectrum. We note that for the fits that we cannot recover the parameters the resulting $\chi^2$-statistic is large, indicating that the systematic errors are dominating the statistical errors. We also note that even for these cases we find the differences in the fitted parameters and the true parameters are at most on the order of a few percent.

For the spectral fitting we find that up to a high exposure time of 4 Ms that we recover the input temperature and the normalization of the spectrum, as well as the isothermal temperature profile. The results of the fit are shown in Table 10 and the temperature profile is shown in Figure 13.


\begin{deluxetable*}{ccccc}
\tabletypesize{\scriptsize}
\tablecaption{Fit to Single Cluster Test, $t_{\rm exp}$ = 240 ks\label{tab8}}
\tablewidth{0pt}
\tablehead{
\colhead{Type} &
\colhead{${r_{\rm c}}$ (kpc)} & 
\colhead{${\beta}$} &
\colhead{${S_0}$ (cts s$^{-1}$cm$^{-2}$arcsec$^{-2}$)} &
\colhead{${\chi^2/{\rm d.o.f.}}$}
}
\startdata
Model & 100 & 1 & $ 8.26 \times 10^{-9}$ & ... \\
Fitted & $100.50^{+3.02}_{-0.50}$ & $1.010^{+0.005}_{-0.021}$ & $(8.18^{+0.11}_{-0.11}) \times 10^{-9}$ & 197.92/195 \\
\enddata
\end{deluxetable*}


\begin{deluxetable*}{cccccccc}
\tabletypesize{\scriptsize}
\tablecaption{Fit to Double Cluster Test, $t_{\rm exp}$ = 240 ks\label{tab9}}
\tablewidth{0pt}
\tablehead{
\colhead{Type} & 
\colhead{${r_{\rm c_1}}$ (kpc)} & 
\colhead{${\beta_1}$} &
\colhead{${S_{0,1}}$ (cts s$^{-1}$cm$^{-2}$arcsec$^{-2}$)} &
\colhead{${r_{\rm c_2}}$ (kpc)} & 
\colhead{${\beta_2}$} &
\colhead{${S_{0,2}}$ (cts s$^{-1}$cm$^{-2}$ arcsec$^{-2}$)} &
\colhead{${\chi^2/{\rm d.o.f.}}$}
}
\startdata
Model & 50 & 1 & $3.78 \times 10^{-8}$ & 200 & 0.7 & $ 1.38 \times 10^{-8}$ & ... \\
Fitted & $47.14^{+2.94}_{-4.35}$ & 1.0 (F) & $(3.90^{+0.27}_{-0.27}) \times 10^{-8}$ & $198.00^{+2.43}_{-6.76}$ & $0.701^{+0.029}_{-0.003}$ & $(1.40^{+0.02}_{-0.02}) \times 10^{-8}$ & 213.93/193 \\  
\enddata
\end{deluxetable*}


\begin{deluxetable}{cccc}
\tabletypesize{\scriptsize}
\tablecaption{Spectral Fit for a Single Cluster, $t_{\rm exp}$ = 4 Ms\label{tab10}}
\tablewidth{0pt}
\tablehead{
\colhead{Type} & 
\colhead{$T_{\rm spec}$ (keV)} & 
\colhead{$N$ ($\frac{10^{-14}}{4\pi\left[D_A(1+z)\right]^2}{\int}n_{\rm e}n_{\rm H}dV$) (cm$^{-5}$)} &
\colhead{$\chi^2/{\rm d.o.f.}$}
}
\startdata
Model & 4.0 & $2.547 \times 10^{-4}$ & ... \\
Fitted & $4.01^{+0.09}_{-0.09}$ & $(2.554^{+0.021}_{-0.021}) \times 10^{-4}$ & 91.40/424 \\
\enddata
\end{deluxetable}


\begin{deluxetable}{ccccc}
\tabletypesize{\scriptsize}
\tablecaption{Systematic Trends in Radial Fits with Increasing Exposure Time\label{tab11}}
\tablewidth{0pt}
\tablehead{
\colhead{$t_{\rm exp}$ (ks)} & 
\colhead{$N_{\rm single}$} &
\colhead{$N_{\rm double}$} &
\colhead{$\chi^2/{\rm d.o.f.}_{\rm single}$} &
\colhead{$\chi^2/{\rm d.o.f.}_{\rm double}$}
}
\startdata
240  & 0 / 3 & 0 / 5 & 1.01 & 1.11 \\
1000 & 0 / 3 & 1 / 5 & 1.07 & 1.20 \\
4000 & 2 / 3 & 5 / 5 & 1.88 & 2.20 \\
\enddata
\end{deluxetable}


\begin{figure}
\plotone{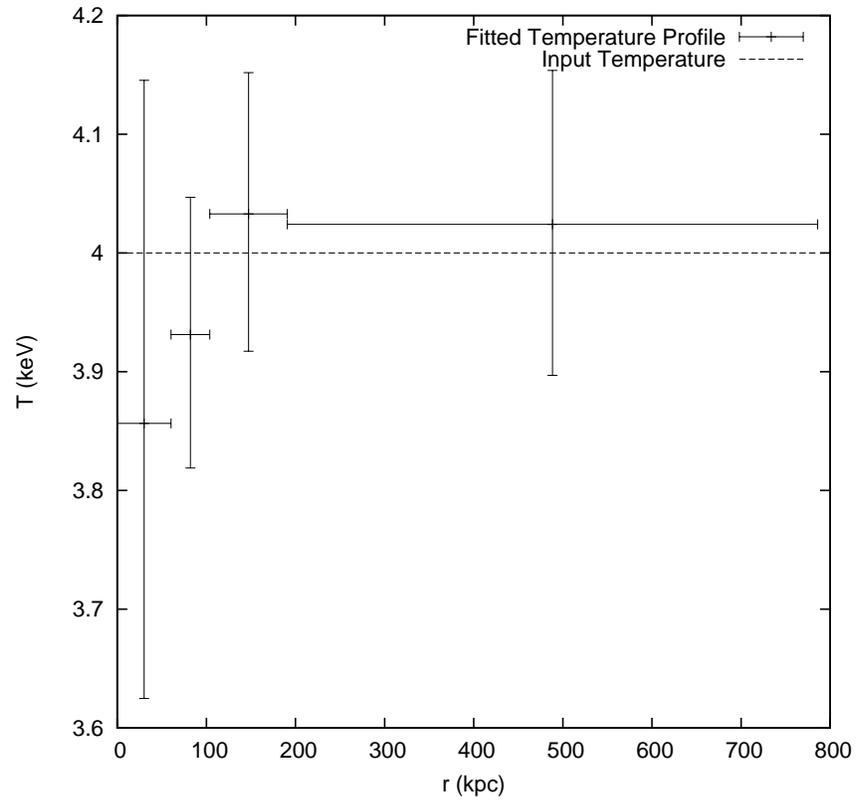}
\caption{Isothermal cluster temperature profile. Dashed line is input temperature.}
\end{figure}

\end{document}